\documentclass[%
 reprint,
 amsmath,amssymb,
 aps,
]{revtex4-1}

\usepackage{graphicx}
\usepackage{dcolumn}
\usepackage{bm}

\begin{document}

\preprint{APS/123-QED}

\title{\boldmath Bootstrap Dynamical Symmetry Breaking with New Heavy Chiral Quarks}

\author{Yukihiro Mimura and Wei-Shu Hou}
 \affiliation{Department of Physics, National Taiwan University, Taipei, Taiwan 10617}

\author{Hiroaki Kohyama
 }
 \affiliation{Department of Physics, Chung-Yuan Christian University, Chung-Li, Taiwan 32023\\
 Department of Physics, Kyungpook National University, Daegu 702-701, Korea}


\begin{abstract}
A Higgs-like new boson with mass around 126 GeV is now established,
but its true nature probably cannot be settled with 2011--2012 LHC data.
We assume it is a dilaton with couplings weaker than the Higgs boson
(except to $\gamma\gamma$ and $gg$),
and explore dynamical symmetry breaking (DSB) by
strong Yukawa coupling of a yet unseen heavy chiral quark doublet $Q$.
Assuming the actual Higgs boson to be heavy, the Goldstone boson
$G$ of electroweak symmetry breaking still couples to $Q$
with Yukawa coupling $\lambda_Q$.
A ``bootstrap" gap equation without a Higgs particle is constructed.
Electroweak symmetry breaking via strong $\lambda_Q$
generates both heavy mass for $Q$,
while self-consistently justifying $G$ as
a massless Goldstone particle in the loop.
The spontaneous breaking of
scale invariance in principle \emph{allows} for a dilaton.
We numerically solve such a gap equation and find
the mass of the heavy quark to be a couple of TeV.
We offer a short critique on the results of
the scale-invariant model of Hung and Xiong,
where a similar gap equation is built with
a massless scalar doublet.
Through this we show that a light SM Higgs at 126 GeV
cannot be viable within our approach to DSB,
while a dilaton with weaker couplings is consistent with our main result.
\begin{description}
\item[PACS numbers]
 14.65.Jk 
 11.30.Qc 
\end{description}
\end{abstract}

\pacs{Valid PACS appear here}
\maketitle


\section{INTRODUCTION and Motivation}

The field of particle physics is in a state of
excitement, accompanied by anxiety.
On one hand, the long-awaited Higgs particle
has finally appeared~\cite{126,125}.
On the other hand, there appears to be
no new physics below the TeV scale,
and one is worried what really stabilizes
the Higgs mass at 126 GeV.

While the existence of a new particle is beyond doubt,
and it is most likely the Higgs boson of the Standard Model (SM),
its true nature needs further scrutiny with more data.
In the July 2012 announcement~\cite{120704}, both the ATLAS and CMS experiments
found enhanced production of the $\gamma\gamma$ mode.
However, despite sensitivity, the CMS experiment
did not seem to detect any signal in the fermionic modes.
These two aspects sparked discussion~\cite{dilaton}
that the object could be the dilaton of scale invariance violation.
The dilaton coupling to vector bosons and fermions are suppressed by $v/f$,
where $v$ is the vacuum expectation value and $f$ the dilaton decay constant.
In contrast, the effective $\gamma\gamma$ and $gg$ couplings
are determined by the trace anomaly, and depend on
the number and nature of extra particles (fermions especially).
Thus, if the observed signal arises mostly from gluon fusion,
a dilaton could mimic the SM Higgs boson.
Discrimination is provided by the detection of
Higgs production through vector boson fusion (VBF),
or bremsstrahlung off a vector boson (VH).
So, in particular, it is the VBF produced $\gamma\gamma$ mode,
and the fermionic $b\bar b$ and $\tau\tau$ modes, that hold the cards.

Both the ATLAS and CMS experiments have updated~\cite{HCP2012}
their results at the High Energy Collider
conference held November 2012 in Kyoto.
However, the updates were uneven in the amount of data used.
Most critically, neither experiments updated the $\gamma\gamma$ result,
while the ATLAS experiment also did not update the $ZZ^{(*)}$ mode.
%
The CMS update for both the $WW^{(*)}$ and $ZZ^{(*)}$ modes
are consistent with SM expectation, but the slightly more
sensitive (to $\mu \equiv$ ratio of measured $\sigma\times{\cal B}$ vs
SM) $WW^{(*)}$ gives $\mu < 1$ by more than one sigma.
For ATLAS, the $WW^{(*)}$ mode was only updated with 2012 data,
giving a result similar to the July result.
The $\mu$ value seems slightly above 1.
For the fermionic final states,
unlike the absence of any hint for $b\bar b$ or $\tau\tau$ in July,
the CMS experiment now sees consistency with $\mu = 1$, i.e. SM expectation.
It is now the ATLAS experiment,
with equivalent amount of data as CMS, that seems to drag a little:
there is no hint of any signal in $b\bar b$,
while $\tau\tau$ is consistent with both SM expection
as well as no signal.
To summarize,
as far as one awaits the $\gamma\gamma$ update,
the $b\bar b$ and $\tau\tau$ modes are not settled~\cite{Tev},
even though things do move towards the SM Higgs.
Since up to 17 to 18 fb$^{-1}$ data has been used,
while a total of $\sim 25$ fb$^{-1}$ is expected for each experiment,
it seems that the fermionic modes cannot be established with
2011-2012 data.

Thus, even though it may appear more and more like the SM Higgs boson,
the dilaton case probably cannot be thrown away offhandedly.
In a general theoretical context~\cite{RattZaff}, it may not be easy to keep
a dilaton much lighter than the scale invariance violation scale.
However, a 126 GeV ``dilaton" would likely remain experimentally allowed.
In this sense, the Achilles' heel to the SM Higgs interpretation
would be the width of the observed particle,
which probably cannot be measured at the LHC;
we probably would not know for decades.
The dilaton width is expected to be $v^2/f^2$ suppressed with respect to
the Higgs width.

Having elucidated the fact that the SM-Higgs nature of
the observed boson cannot be fully established,
and conversely the dilaton possibility cannot be firmly ruled out,
in this paper we wish to explore a scenario of strong-coupling
driven dynamical symmetry breaking (DSB) that allows for the dilaton.
Assuming the actual Higgs boson is very heavy or doesn't even exist,
let us check what is the firm experimental knowledge.
The Higgs mechanism is an experimental fact.
That is, the electroweak (EW) gauge symmetry is
experimentally established, while the gauge bosons,
as well as the chirally charged fermions, are all
found to be massive. Thus, the Goldstone particle
of electroweak symmetry breaking (EWSB) is ``eaten"
by the EW gauge bosons, which become massive (the Meissner effect),
as has been experimentally established since 30 years.

With both quarks and gauge bosons massive,
without invoking the Higgs boson explicitly,
one can see heuristically~\cite{Hou12} that,
starting from the left-handed gauge coupling,
the longitudinal component of the EW gauge bosons,
i.e. equivalently the Goldstone bosons,
couple to quarks by the standard Yukawa coupling.
Thus, Yukawa couplings are experimentally established.
Furthermore~\cite{Hou12},
much of flavor physics and $CP$ violation
studies probe the effects of Yukawa couplings,
providing ample support for their existence.

\begin{figure}[b!]
\begin{center}
\vskip-0.3cm
 \includegraphics[width=88mm]{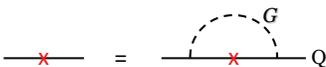}
\vskip-5.1cm
\caption{``Bootstrap" gap equation for generating
heavy quark mass from Goldstone boson loop.}
 \label{GapEq}
\end{center}
\end{figure}

With the existence of three quark generations already,
interest~\cite{4S4G} has been growing in the fourth generation
in the past few years, as the LHC opens up new horizons.
Of course, an SM Higgs interpretation of the 126 GeV
boson deflates one's faith in the 4th generation.
However, with our assumption of a heavy Higgs boson
while taking the 126 GeV boson as the dilaton,
the 4th generation remains a viable possibility.
The pursuit has indeed been vigorous at the LHC
in the past few years, where the current bound~\cite{bpCMS12,leptCMS12}
on $m_Q$ has reached beyond 600 GeV
(we shall assume ``heavy isospin" symmetry, $I_Q$,
and treat the doublet $Q$ as degenerate).
This is already above the nominal perturbative
partial wave unitarity bound (UB) of 550 GeV~\cite{Chanowitz78}.
The Yukawa coupling $\lambda_Q \equiv \sqrt2 m_Q/v$
is more than 3.5 times larger than $\lambda_t \simeq 1$,
and has entered the strong coupling regime
($\alpha_Q \equiv \lambda_Q^2/4\pi \sim 1$).

With the ever increasing bound on $m_Q$,
it may well not exist. But being beyond UB,
it begs the question:
Could the \emph{strong Yukawa coupling} of
$Q$ generate~\cite{Holdom06} EWSB itself?
Along this line, a gap equation,
given symbolically in Fig.~1,
was constructed~\cite{Hou12} without ever invoking
the Higgs doublet, or the Higgs boson field.
The logic, or philosophy goes as follows.
The Goldstone boson $G$ is viewed as
a tightly bound (by Yukawa coupling itself!)
$Q\bar Q$ state. It was in fact \emph{postulated}~\cite{Hou12}
as the collapsed state by Yukawa binding,
as seen through a Bethe--Salpeter equation study~\cite{Jain}
(for further elucidation, see Ref.~\cite{EHY11}),
which is taken as suggestive of triggering EWSB itself.
With no New Physics in sight at the LHC,
not even the heavy chiral quark $Q$ itself,
the loop momentum integration runs up to roughly $2m_Q$,
without the need to add any further effects
(with the simplification of ignoring, or truncating,
corrections to $G$ propagation and $GQQ$ vertex).
This is therefore a ``bootstrap" gap equation,
in that the strong Yukawa coupling itself is
the source of EWSB, or mass generation for quark $Q$,
which simultaneously justifies keeping
the Goldstone $G$ in the loop.
The \emph{existence} of a large Yukawa coupling $\lambda_Q$
is used as input, \emph{without a theory} for $\lambda_Q$ itself.

The question now is whether one could find a
solution to such a gap equation.
If so, we would have demonstrated the case for DSB,
and the potential riches that could follow.
The purpose of this paper is to
formulate more clearly the gap equation,
and demonstrate that numerical solution does exist
at strong coupling.
A similar gap equation was formulated by
Hung and Xiong~\cite{HX11}, where $G$ in Fig.~1 is
replaced by a massless Higgs doublet field.
We will compare and offer a critique.

The paper is organized as follows.
In the next section,
after briefly mentioning the Nambu--Jona-Lasinio model,
which is a simplified template for DSB where the self-energy
is momentum-independent,
we focus on the more relevant strongly-coupled scale-invariant QED,
which we utilize as a means of setting up our approach.
In Sec. III, we formulate our bootstrap gap equation
following the setup in Sec. II, and solve numerically
for the critical $\lambda_Q$ (hence $m_Q$).
We compare with a similar study by Hung and Xiong,
and pursue the consequence of taking the physical loop cutoff
to be less than $2m_Q$, and outline the framework for further study.
In Sec. IV we discuss the many questions and issues of DSB
related to our bootstrap equation, including the issue of the dilaton.
Our conclusion is given in Sec. V.


\section{Historical Backdrop for DSB}

In this section we briefly review
the Nambu--Jona-Lasinio model,
where one sets up a gap equation with
its well-known solution.
We then turn to the so-called
strongly-coupled scale-invariant QED,
which is closer to our gap equation.
By recounting some major steps, we also set up
our notation for later usage.

\subsection{ NJL Model }

The Nambu--Jona-Lasinio model~\cite{NJL}, proposed in 1960,
is the earliest, explicit model of DSB,
where the breaking of global chiral symmetry leads to
generation of nucleon mass, and
the pion as a (pseudo-)Goldstone boson~\cite{NG}.

\begin{figure}[t!]
\begin{center}
 \includegraphics[width=88mm]{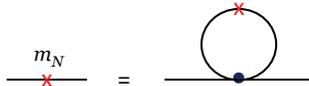}
\vskip-5cm
\caption{Gap equation for the Nambu--Jona-Lasinio model
for generating nucleon mass.}
 \label{GapNJL}
\end{center}
\end{figure}

The model can be depicted as in Fig.~2, where a
four-fermion interaction is introduced,
represented by the blob on the right-hand side.
The nucleon mass, represented by a cross,
is self-consistently generated.
One easily finds the gap equation
\begin{eqnarray}
 m_N &=& \frac{N_C}{8\pi^2}G\int_0^{\Lambda^2} dq^2\, q^2 \frac{m_N}{q^2 + m_N^2} \nonumber \\
     &=& \frac{N_C}{8\pi^2}G\Lambda^2
         \left(1 - \frac{m_N^2}{\Lambda^2} \log\left(1 + \frac{\Lambda^2}{m_N^2}\right)
         \right)m_N,
 \label{mN}
\end{eqnarray}
where $G$ here is the four-fermi coupling and $\Lambda$ is the cutoff.
Since $m_N$ on both sides factor out, one has
\begin{eqnarray}
 1 - \frac{G_{\rm crit}}{G} &=& \frac{m_N^2}{\Lambda^2} \log\left(1 + \frac{\Lambda^2}{m_N^2}\right),
 \label{NJL}
\end{eqnarray}
which admits a solution for $G > G_{\rm crit}$, where
\begin{eqnarray}
 {G_{\rm crit}} &=& \frac{8\pi^2}{N_C \Lambda^2}.
 \label{Gcrit}
\end{eqnarray}

To understand what is happening, one can iterate the cross
of the left-hand side of Fig.~2 on the right-hand side,
and sees that it contain an infinite number of diagrams.
This effectively puts the original self-energy diagram
into the denominator, and in the end, one trades
the parameters $G$ and $\Lambda$ for the
physical $m_N$ and the pion-nucleon coupling.
At the more refined level and
using the quark language, one can show further that
the emergent Goldstone boson, the pion, is in fact
a ladder sum of the quark-level four-fermi interaction.

We will return at the end to discuss the similarity
and differences of the NJL model with our gap equation.


\subsection{ Strongly-coupled Scale-invariant QED }


We note that the self-energy bubble of Fig.~2 does not depend on
external momentum $p$, so at the superficial level,
Fig.~2 is quite different from Fig.~1.
We now turn to QED, where there is closer similarity.


The general gap equation for QED~\cite{FK76} can be written in the form
of the Schwinger--Dyson (SD) equation,
\begin{eqnarray}
\Sigma(p) = ie^2 \int \frac{d^4 q}{(2\pi)^4}
\gamma^\mu D_{\mu\nu} (q) S(p-q) \Gamma^\nu (p,q),
 \label{SD-QED}
\end{eqnarray}
where $\Sigma = /\!\!\!p - m - S(p)^{-1}$ is the electron self-energy
 with $S$ the electron (full) propagator,
 $D_{\mu\nu}$ is the photon (full) propagator,
 and $\Gamma$ is the full vertex.

\subsubsection{ Ladder Approximation and Integral Form }

Truncating the exact, full vertex and photon propagator
by the approximation,
\begin{eqnarray}
\Gamma^\nu(p,q) &=& \gamma^\nu, \\
D_{\mu\nu}(q) &=&
\frac{-g_{\mu\nu} + q^\mu q^\nu/q^2}{q^2} - \xi \frac{q_\mu q_\nu}{q^4},
 \label{Dmunu}
\end{eqnarray}
which is called the ladder (or rainbow) approximation,
the gap equation becomes
\begin{equation}
S(p)^{-1} = /\!\!\!p
           - i e^2 \int \frac{d^4q}{(2\pi)^4} \gamma^\mu D_{\mu\nu}(p-q) S(q) \gamma^\nu,
 \label{gapQED}
\end{equation}
with $D_{\mu\nu}$ given in Eq.~(\ref{Dmunu}),
and we have set $m = 0$, i.e. massless QED (at Lagrangian level).
Pictorially, this is represented as in Fig.~3.
The question now is whether the scale invariance could be
dynamically broken at some
strong coupling $\alpha = e^2/4\pi$?

\begin{figure}[t!]
\begin{center}
\vskip-2cm
 \includegraphics[width=80mm]{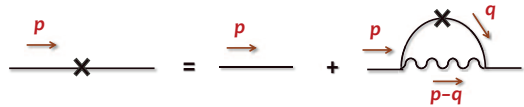}
\vskip-2.1cm
\caption{Gap equation for QED in the ladder approximation.}
 \label{GapQED}
\end{center}
\end{figure}

We define the electron propagator as~\cite{FK76}
\begin{equation}
S(p) = \frac{1}{A(p^2)\, /\!\!\!p - B(p^2)},
 \label{Se}
\end{equation}
where the $A$ term corresponds to wave function renormalization,
i.e. related to the usual $Z^{-1}$ factor.
A finite pole of the propagator, $p^2 = B^2(p^2)/A^2(p^2)$,
would give the dynamical effective mass.
Our aim is therefore solving $A$ and $B$ from the gap equation
of Eq.~(\ref{gapQED}). Inserting Eq.~(\ref{Dmunu})
and after some algebra, one finds
\begin{equation}
B(p^2) =  e^2 \int \frac{id^4q}{(2\pi)^4}
\left(\frac{-4+1-\xi}{k^2} \right)
\frac{B(q^2)}{A^2(q^2) q^2 - B^2(q^2)},
\end{equation}
and
\begin{equation}
A(p^2) = 1 - \frac{e^2}{4p^2}
{\rm tr}
\int \frac{id^4q}{(2\pi)^4}
\gamma^\mu D_{\mu\nu}^{(\xi)}
\frac{A(q^2) /\!\!\!q}{A^2(q^2) q^2 - B^2(q^2)} \gamma^\nu /\!\!\!p.
\end{equation}

\subsubsection{ Landau Gauge and Differential Form }

Simplification can be achieved in Landau gauge, $\xi = 0$,
where one finds $A(p^2) = 1$. Since $A$ is the inverse of
wave function renormalization, this means that it satisfies
Ward-Takahashi identity even under the ladder approximation.
The gap equation we need to solve is simplified to just
an equation for $B$,
\begin{equation}
B(p^2) =  -3 e^2 \int
\frac{id^4q}{(2\pi)^4} \frac{B(q^2)}{(p-q)^2 (q^2- B^2(q^2))}.
\end{equation}
After Wick rotation, and using
%
\begin{equation}
\int d^4q\frac{f(q^2)}{(p-q)^2} =
\pi^2 \int dq^2 q^2 \left[\frac{\theta(p-q) }{p^2}+\frac{\theta(q-p) }{q^2}\right]f(q^2),
 \label{app_step}
\end{equation}
one obtains
\begin{equation}
B(x) = \frac{3\alpha}{4\pi}
\left(
\frac1{x} \int_{\Lambda_{\rm IR}^2}^x dy \frac{y B(y)}{y+ B^2(y)}
+ \int_x^{\Lambda^2} dy \frac{B(y)}{y+B^2(y)}
\right),
 \label{Bqed}
\end{equation}
where $x = p_E^2$ is the Euclidean momentum squared, and
$\Lambda^2$, $\Lambda_{\rm IR}^2$ are its ultraviolet (UV) and infrared (IR) cutoffs, respectively.

The integral equation can be changed to a differential equation
by noting
\begin{eqnarray}
\frac{dB(x)}{dx}
&=& \frac{3\alpha}{4\pi}
\left(-\frac{1}{x^2}\right)\int_{\Lambda_{\rm IR}^2}^x dy \frac{y B(y)}{y+ B^2(y)}, \\
 \label{dBdx}
%
%
\frac{d(xB(x))}{dx} &=& \frac{3\alpha}{4\pi}
\int_x^{\Lambda^2} dy \frac{B(y)}{y+B^2(y)}.
\end{eqnarray}
One obtains the differential equation,
\begin{equation}
x \frac{d^2B(x)}{dx^2} + 2 \frac{dB(x)}{dx}
+ \frac{3\alpha}{4\pi} \frac{B(x)}{x+B^2(x)} = 0.
 \label{diffBqed}
\end{equation}
together with the boundary conditions (B.C.) for IR and UV,
which are given as
\begin{eqnarray}
\left.\frac{dB(x)}{dx}\right|_{x=\Lambda_{\rm IR}^2} &=& 0, \ \ \ \
\left.\frac{d(xB(x))}{dx}\right|_{x=\Lambda^2} = 0.
\end{eqnarray}
If IR cutoff is 0, the B.C. for IR should be replaced by
$2BdB/dx = - 3\alpha/4\pi$ at $x=0$.

\subsubsection{ Analytic Solution in Landau Gauge }

In order to study qualitative features,
let us find first an approximate solution.
In particular, for the special range for the IR cutoff,
let us take $B(x) \ll \Lambda_{\rm IR} < x^{\frac12} < \Lambda$,
then Eq.~(\ref{diffBqed}) is simplified to
\begin{equation}
x^2 \frac{d^2B(x)}{dx^2} + 2 x\frac{dB(x)}{dx}
+ \frac{3\alpha}{4\pi} {B(x)} = 0.
 \label{diffBqed0}
\end{equation}
The characteristic equation for
solution $B(x) = C x^\lambda$ is
\begin{equation}
\lambda (\lambda-1) + 2\lambda + \frac{3\alpha}{4\pi}=0,
\end{equation}
with discriminant $1- 3\alpha/\pi$, hence
the behavior is different for $\alpha > \alpha_c$ and $\alpha < \alpha_c$,
where $\alpha_c = \pi/3$.
%
The analytical solution under the approximation is given as
\begin{eqnarray}
B &=& b_1\, x^{(-1+\sqrt{1-\alpha/\alpha_c})/2} + b_2\, x^{(-1-\sqrt{1-\alpha/\alpha_c})/2}
 \nonumber \\
 &\equiv& b_1\, x^{c_1} + b_2\, x^{c_2},
 \label{Banal}
\end{eqnarray}
and the boundary conditions can be written as
\begin{eqnarray}
\left(
 \begin{array}{cc}
  c_2 & c_1 \Lambda_{\rm IR}^{2(c_1-c_2)} \\
  -1 & \Lambda^{2(c_1-c_2)}
 \end{array}
\right)
\left(
 \begin{array}{c}
  b_1 \\ b_2
 \end{array}
\right)=
\left(
 \begin{array}{c}
  0 \\ 0
 \end{array}
\right).
 \label{b1b2}
\end{eqnarray}
The determinant must vanish for nontrivial $b_1, b_2$, hence
\begin{equation}
\frac{\sqrt{1-\alpha/\alpha_c}-1}{\sqrt{1-\alpha/\alpha_c}+1} =
\left(\frac{\Lambda^2}{\Lambda_{\rm IR}^2}\right)^{\sqrt{1-\alpha/\alpha_c}}.
 \label{det0}
\end{equation}
To satisfy this condition, $\alpha> \alpha_c = \pi/3$ is needed,
and for given $\Lambda/\Lambda_{\rm IR}$,
$\alpha$ takes on discontinuous values,
\begin{equation}
n\pi + \tan^{-1}\frac1{\sqrt{\alpha/\alpha_c-1}} = \sqrt{\alpha/\alpha_c-1}
 \log \frac{\Lambda}{\Lambda_{\rm IR}}.
\end{equation}
One sees that $\alpha \to \alpha_c$ for
$\Lambda/\Lambda_{\rm IR} \to \infty$,
the nominal ``continuum limit".
For $\alpha < \alpha_c$,
the only solution that satisfies the B.C.
is the trivial $B(x) = 0$.
%

%
%
%

The approximate solution for small IR cutoff
can now be obtained by replacing $B(x) = m$ (constant) for small $x$.
The differential equation becomes
\begin{equation}
x \frac{d^2B(x)}{dx^2} + 2 \frac{dB(x)}{dx}
+ \frac{3\alpha}{4\pi} \frac{B(x)}{x+m^2} = 0,
 \label{Hyper}
\end{equation}
where $B(x)$ is a hypergeometric function.
%
%
%
%
%
Namely,
\begin{equation}
B(x) = m\, F\left(\frac12 - i\gamma,\; \frac12 + i\gamma,\; 2;\ -\frac{x}{m^2}\right),
\end{equation}
where
\begin{equation}
i\gamma = \frac12 \sqrt{1-\frac{\alpha}{\alpha_c}},
\qquad
\alpha_c = \frac{\pi}{3},
\end{equation}
and $F$ is a hypergeometric function.
Checking the asymptotic behavior for $x\gg m^2$,
the power behavior for $\alpha < \alpha_c$
cannot satisfy the UV boundary condition.

The gap equation has a nontrivial or oscillatory solution
for $\alpha > \alpha_c$, where the critical coupling
is $\alpha_c = \pi/3$ for QED.
The dynamical effective mass is obtained by
solving $x = B^2(-x)$.
Note that the pole of the propagator should be given in
the time-like region, so one has to make
analytical continuation in order to obtain a physical mass,
which can be done smoothly for $\Lambda_{\rm IR} \to 0$.

An issue arises in that the dynamical mass
\begin{equation}
m = 4\Lambda\, e^{-\frac{\pi}{\sqrt{\alpha/\alpha_c-1}}}.
\end{equation}
is proportional to the UV cutoff $\Lambda$.
For $m$ to be physical, however, it should not depend on $\Lambda$.
Miransky suggested~\cite{Miransky} that $\alpha \to \alpha_c$ as $\Lambda \to \infty$,
i.e. $\alpha_c$ is a nontrivial UV fixed point.
A related issue, which we would not go into, is
whether there would be a dilaton associated
with breaking of scale invariance~\cite{dilaton1}.


\section{Bootstrap Dynamical EWSB}

For the gap equation~\cite{Hou12} with large empirical
Yukawa coupling, one could continue to follow the
Fukuda--Kugo approach of the previous section.
In Landau gauge ($\xi = 0$),
the propagation of (would-be) Goldstone modes and gauge bosons
are properly separated,
and the gap equation becomes equivalent to our discussion below
if one takes the $g\to 0$ limit for the gauge coupling.
The Goldstone $G^\pm$ and $G^0$ couple to
fermions with the familiar Yukawa couplings,
which we have argued~\cite{Hou12} as experimentally established.
They are also unaltered by the $g\to 0$ limit.
The main assumption is the addition of a
new (heavy) chiral quark doublet,
where the $I_Q$ heavy isospin implies that
$\lambda_U = \lambda_D \equiv \lambda_Q$, i.e. equality of
the $I_{Q3} = \pm \frac12$ Yukawa couplings.
We ask whether large $\lambda_Q$ could be the source of EWSB,
through the conceptual gap equation of Fig.~1.

\subsection{ Ladder Approximation }

\begin{figure}[t!]
\begin{center}
\vskip0.4cm
 \includegraphics[width=80mm]{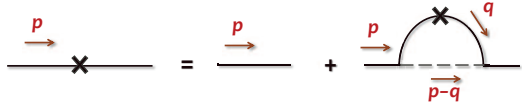}
\vskip-4.5cm
\caption{Gap equation for large Yukawa coupling
         in the ladder approximation,
         where $m_0$ vanishes by gauge (or chiral) invariance.}
 \label{GapYuk}
\end{center}
\end{figure}

In any case, we do not know the full propagator and vertex functions.
We approximate the Goldstone-fermion vertex as undressed,
and the Goldstone propagator remains as $i/k^2$,
analogous to the QED treatment in previous section.
Similar to Fig.~\ref{GapQED},
the gap equation for large Yukawa (vanishing $g$) coupling
is depicted in Fig.~\ref{GapYuk}.
It is interesting to note that, while $m_0 = 0$ for strong QED
was set by hand, in the present case, $m_0 = 0$ is by
gauge (or chiral, in $g \to 0$ limit) invariance.

We aim at solving for quark propagator $S(p)$,
given as
\begin{equation}
S(p)^{-1} = A(p^2)\, /\!\!\!p - B(p^2),
 \label{SQ-1}
\end{equation}
which is in same form as Eq.~(\ref{Se}) for QED.
Following similar steps as in Sec.~II, and
assuming $\lambda_U = \lambda_D = \lambda_Q$ degeneracy,
we obtain
\begin{eqnarray}
B_{p^2} &=\frac{3\lambda_Q^2}{2}&
\int \frac{d^4q}{i(2\pi)^4}
\frac{1}{(p-q)^2}
\frac{
B_{q^2}
 }{q^2 A^2_{q^2} - B^2_{q^2}}\nonumber \\
&-\frac{\lambda_Q^2}2&
 \int \frac{d^4 q}{i(2\pi)^4}
\frac{1}{(p-q)^2-m_h^2}
\frac{
B_{q^2}
 }{q^2 A^2_{q^2} - B^2_{q^2}},
 \label{Bp2}
\end{eqnarray}
and
\begin{eqnarray}
A_{p^2} = 1 + &\frac{3\lambda_Q^2}{2p^2}&
\int \frac{d^4q}{i(2\pi)^4}
\frac{p\cdot q}{(p-q)^2}
\frac{
A_{q^2}
 }{q^2 A^2_{q^2} - B^2_{q^2}}\nonumber \\
+ &\frac{\lambda_Q^2}{2p^2}&
 \int \frac{d^4 q}{i(2\pi)^4}
\frac{p\cdot q}{(p-q)^2-m_h^2}
\frac{
 A_{q^2}
 }{q^2 A^2_{q^2} - B^2_{q^2}},
 \label{p2Ap2}
\end{eqnarray}
where placement of $i$ anticipates the Wick rotation,
and $B_{q^2}$ stands as shorthand for $B(q^2)$,
and likewise for $A$.
We have already used $\xi = 0$,
so, compared to massless QED, we now have to consider $A$,
or wave function renormalization effects,
hence a coupled set of equations for $A(p^2)$ and $B(p^2)$.
Note that we have kept a ``Higgs" term, applying
Standard Model Higgs boson, $h^0$, couplings.
This is for purpose of later comparison with
the work of Hung and Xiong~\cite{HX11}.

If one simply drops the second term (no physical Higgs,
or taking $m_h \sim \infty$), so only Goldstone modes
propagate, the gap equation becomes,
after angular integration and Wick rotation,
\begin{eqnarray}
B(x) &=& \kappa_b
\left(
\frac1{x} \int_{\Lambda_{\rm IR}^2}^x dy \frac{y B(y)}{y A^2(y)+ B^2(y)} \right. \nonumber \\
 && \left. \quad\;\ \
 + \int_x^{\Lambda^2} dy \frac{B(y)}{y A^2(y)+B^2(y)} \right),
 \label{Bx} \\
A(x) = 1 &+& \kappa_a
\left(
\frac1{x^2} \int_{\Lambda_{\rm IR}^2}^x dy \frac{y^2 A(y)}{y A^2(y)+ B^2(y)} \right. \nonumber \\
 && \left. \quad\quad\;
+ \int_x^{\Lambda^2} dy \frac{A(y)}{y A^2(y)+B^2(y)}
\right),
 \label{Ax}
\end{eqnarray}
where
\begin{equation}
\kappa_b = 2\kappa_a = \frac{\frac32\lambda_Q^2}{16\pi^2}
 = \frac{\frac32\alpha_Q}{4\pi}. \quad\quad {\rm (no\ Higgs)}
 \label{noH}
\end{equation}
Had we taken the limit $m_h \to 0$ in Eqs.~(\ref{Bp2}) and (\ref{p2Ap2})
to mimic the Hung-Xiong approach of a massless scalar doublet, then
\begin{equation}
\kappa_b = \kappa_a = \frac{\lambda_Q^2}{16\pi^2}
 = \frac{\alpha_Q}{4\pi}. \quad\ {\rm (massless\ Higgs)}
 \label{mh0}
\end{equation}
We keep the notation of $\kappa_b$ and $\kappa_a$ in
Eqs. (\ref{noH}) and (\ref{mh0}) to cover these two cases.


Analogous to Sec.~II, Eqs.~(\ref{Bx}) and (\ref{Ax}) can be
put into differential form,
\begin{eqnarray}
&&x B^{\prime\prime} + 2 B^\prime + \frac{\kappa_bB}{x A^2 + B^2} = 0 ,
 \label{xBpp} \\
&&x A^{\prime\prime} + 3 A^\prime + \frac{2\kappa_aA}{x A^2+B^2} =0,
 \label{xApp}
\end{eqnarray}
with the boundary conditions
\begin{eqnarray}
\left.B^\prime(x)\right|_{x=\Lambda^2_{\rm IR}} &=& 0,
\  \left(xB^\prime(x)+B(x)\right)|_{x=\Lambda^2} = 0, \\
 \left.A^\prime(x)\right|_{x=\Lambda^2_{\rm IR}} &=& 0,\
\left(\frac12x A^\prime(x)+ A(x)\right)|_{x=\Lambda^2} = 1,
\end{eqnarray}
where prime stands for $x$-derivative.

We note at this point that, if one ignores wave function renormalization,
i.e. Eq.~(\ref{xApp}), while forcefully setting $A(x)=1$ in Eq.~(\ref{xBpp}),
then one has the same solution as in QED,
with the change in critical coupling
\begin{equation}
\alpha_Q^c = \frac{2\pi}{3},
\end{equation}
which is twice as high as for QED, and
\begin{equation}
\lambda^c_Q = 2\pi\sqrt{\frac23} \simeq 5.13,
 \label{lamQED}
\end{equation}
hence superficially a ``critical mass" $m_Q^c \simeq 890$ GeV,
which is above current LHC limits~\cite{bpCMS12,leptCMS12}.
But it should be clear that the wave function renormalization
$A(p^2)$ term cannot be neglected~\cite{foot}.

\subsection{ Numerical Solution }


Redefining $p^2 = x = e^{2t}$,
our target simultaneous differential equations with B.C. become
\begin{eqnarray}
&&\ddot{B} + 2 \dot{B} + \frac{4\kappa_bB}{A^2+B^2 e^{-2t}}=0, \\
&&\ddot{A} + 4 \dot{A} + \frac{8\kappa_aA}{A^2+B^2 e^{-2t}}=0, \\
&&\dot{B}(t_{\rm IR}) = 0, \;\ \quad \dot{B}(t_{\rm UV}) + B(t_{\rm UV}) =0, \\
&& \dot{A} (t_{\rm IR}) = 0,\quad \frac14\dot{A}(t_{\rm UV}) + A(t_{\rm UV}) = 1,
\end{eqnarray}
where dot represents $t$-derivative,
and $e^{t_{\rm UV}} = \Lambda_{\rm UV} = \Lambda$,
and $e^{t_{\rm IR}} = \Lambda_{\rm IR}$.

   \subsubsection{ Asymptotic Properties }

Due to scale invariance,
the differential equations are invariant under
\begin{eqnarray}
&&x \to a^2 x   \qquad (t \to t+ \log a), \\
&&\Lambda_{\rm UV,IR} \to a \Lambda_{\rm UV,IR}, \\
&&B \to a B, \quad A\to A.
\end{eqnarray}
As a result, the solutions of the differential equations depend only on
$\Lambda_{\rm UV}/\Lambda_{\rm IR}$ ($=e^{t_{\rm UV}- t_{\rm IR}}$) and
$m_{\rm dyn} \equiv B(t_{\rm IR})/A(t_{\rm IR})$
for given $\kappa_a$ and $\kappa_b$.
Thus, $m_{\rm dyn}$ is a kind of integration constant.
We will see that the most important feature of the solutions is that
only special values (discontinuous values) of $\kappa_a$ and $\kappa_b$
are allowed for given boundary conditions.

If we take $\kappa_a=0$, the equations can be solved analytically,
and the solution can be described as in the case of
strong coupling QED with Landau gauge,
Eqs.~(\ref{Banal})--(\ref{det0}).
The property illustrated for $\kappa_a = 0$
should hold also for $\kappa_a \neq 0$,
where, depending on $\Lambda_{\rm UV}/\Lambda_{\rm IR}$,
$\kappa_a$ should take on special discontinuous values
to satisfy the gap equation for the
cases of $\kappa_a/\kappa_b = 1$ or $1/2$.

To see this, we note that the
$B^2 e^{-2t}$ term in the denominators are irrelevant.
This is because for $t \ll \log m$,
the $t$ dependence of $A$ and $B$ are negligible due to
the boundary conditions $\dot A(t_{\rm IR}) = \dot B(t_{\rm IR})=0$.
Therefore, to understand the behavior of the solution,
one can take $t_{\rm IR} \sim \log m$ and consider
the differential equation with $B^2 e^{-2t}$ dropped:
\begin{eqnarray}
&&\ddot{B} + 2 \dot{B} + \frac{4\kappa_bB}{A^2}=0, \\
&&\ddot{A} + 4 \dot{A} + \frac{8\kappa_a}{A}=0,
\end{eqnarray}
where the equation for $A$ becomes independent from $B$,
but its solution affects $B$.
%
%
%
%

The solution of $A$ is obtained analytically:
\begin{eqnarray}
A(x) &=& A_0 \frac{\Lambda_{\rm UV}^4}{x^2} e^{-\zeta^2} \sqrt{\frac{\kappa_a}{\pi}} , \\
\zeta &=& {\rm Erf}^{-1} \left[\frac{1}{A_0} \left(1- \frac{x^2}{\Lambda_{\rm UV}^4}\right) +
 {\rm Erf}\left(\frac{1}{\sqrt{\kappa_a}}\right)\right],
\end{eqnarray}
where $A_0$ is an integration constant which can be fixed by the B.C.
for IR.
Using the analytical solution, one can show
that only discontinuous values of $\kappa_{a,b}$
can satisfy the B.Cs.
The ``critical" value of $\kappa_b$ can be
easily obtained numerically, even without neglecting $B^2 e^{-2t}$ term
in the denominators.
Of course, since the solutions are found numerically,
it is not a proof that one really has a ``critical" value.
The upshot is that only special values of the coupling
can satisfy the SD equation for given values of
$\Lambda_{\rm UV}/\Lambda_{\rm IR}$ for $m_{\rm dyn} < \Lambda_{\rm IR}$
(or $\Lambda_{\rm UV}/m_{\rm dyn}$ for $m_{\rm dyn} > \Lambda_{\rm IR}$),
where $m_{\rm dyn} = B(\Lambda_{\rm IR})/A(\Lambda_{\rm IR})$.

Our numerical solution gives  
\begin{eqnarray}
&&\kappa_b^c \simeq 1.4   \;\, \qquad (\kappa_b = 2\kappa_a = 3\alpha_Q/8\pi), \\
&&\kappa_b^c \simeq 13.7   \qquad (\kappa_b = \kappa_a = \alpha_Q/4\pi).
\end{eqnarray}
corresponding to
\begin{eqnarray}
&&\lambda_Q^c \simeq 12  \qquad (\kappa_b = 2\kappa_a = 3\alpha_Q/8\pi),
 \label{lamcr_infty} \\
&&\lambda_Q^c \simeq 46  \qquad (\kappa_b = \kappa_a = \alpha_Q/4\pi),
 \label{lamcrHX_infty}
\end{eqnarray}
where the latter case is much higher.
Here $c$ stands for ``critical", and our
numerical values are extracted in
the large $\Lambda_{\rm UV}/\Lambda_{\rm IR}$
and $\Lambda_{\rm UV}/m_{\rm dyn}$ limit.
Note that for the case of $\kappa_a = 0$ (i.e. $A = 1$),
the critical value was $\lambda_Q^c \simeq 5.1$,
Eq.~(\ref{lamQED}).


The values in Eqs.~(\ref{lamcr_infty}) and (\ref{lamcrHX_infty})
corresponds to effectively taking $\Lambda_{\rm UV} \to \infty$,
which is certainly not the range of validity
for Eq.~(\ref{lamcr_infty}) as a descendent of Fig.~1.
That is, the conceptual foundation for Fig.~1 is that,
for momentum roughly up to somewhere below $2m_Q$,
corrections to the Goldstone boson propagator and vertex
has been ignored.
Nevertheless, at face value,
if we naively apply the physical $v \simeq 246$ GeV,
then Eqs.~(\ref{lamcr_infty}) and (\ref{lamcrHX_infty})
imply the mass values
\begin{equation}
 m_Q^c > 2.1\ {\rm TeV}, \quad\quad {\rm (No\ Higgs)}
 \label{mQcr_infty}
\end{equation}
and 8.1 TeV, respectively, which are rather high.
We no longer display the latter, ``massless Higgs"
(or light Higgs) case, not only because it is
clearly out of reach for the LHC,
but for theoretical reasons as we explain below.
The lower bound nature of Eq.~(\ref{mQcr_infty})
would be explained subsequently.


\subsubsection{ Comparison with work of Hung and Xiong }


We now make some comparison with, and offer a critique of,
the approach of Hung and Xiong~\cite{HX11} (HX).
We have mimicked the concept of HX by keeping a
``Higgs" scalar contribution in Eqs.~(\ref{Bp2}) and (\ref{p2Ap2}),
which resulted in the second case of Eq.~(\ref{mh0})
in the limit of $m_h \to 0$, as compared with
our case of interest, Eq.~(\ref{noH}),
where we drop the $m_h$-dependent term.

HX \emph{assumed} the existence of a massless Higgs doublet,
where our Eqs.~(\ref{xBpp}) and (\ref{xApp})
with Eq.~(\ref{mh0}) should be a faithful representation.
However, not only is the massless doublet assumed,
HX ignored wave function renormalization, i.e. the $A$ term,
in the treatment of their gap equation.
Furthermore, the sign of our second integral in Eq.~(\ref{Bp2})
disagrees with HX,
suppressing the coefficient of $\lambda_Q^2$ in comparison.
In the earlier work~\cite{Hou12} that set up
the gap equation of Fig.~1, taking the numerics of HX,
the estimate of $\lambda_Q^c = \sqrt2 \pi$ (compare Eq.~(\ref{lamQED}))
gave $m_Q^c \simeq 770$ GeV,
which is not too far above the current LHC bound.
But with our sign for the second integral in Eq.~(\ref{Bp2}),
we would arrive at $\lambda_Q^c = 2 \pi$, or $m_Q^c \sim 1.1$ TeV,
i.e. a factor of $\sqrt2$ higher.
With our results, the Higgs boson effect
would cancel out part of the Goldstone effect,
hence requiring stronger Yukawa coupling.

But we have argued that it is not justified to ignore
the wave function renormalization effect of $A(p^2)$.
After all, the boson loop has momentum dependence,
so it would necessarily affect the $Z$ factor.
Thus, the above simple numerics is incorrect.
Keeping $A$ in our numerical study,
hence the coupled $A$--$B$ equations,
a considerably higher critical $m_Q^c$ is found.
For the case of taking
$\kappa_b = \kappa_a = \lambda_Q^2/16\pi^2$, Eq.~(\ref{mh0}),
where we mimic HX's massless Higgs doublet effect,
the critical $\lambda_Q^c$ of Eq.~(\ref{lamcrHX_infty})
is almost 4 times as high as that for Eq.~(\ref{lamcr_infty}).
In fact, we obtain the $A(\Lambda_{\rm UV}^2) = 2.7$,
which is quite different from 1.

Our criticism goes far deeper.
Taking a scalar doublet as massless,
such that superficially one has ``scale-invariance",
is totally \emph{ad hoc}.
Effectively one has to \emph{hold} the parameters
of the Higgs potential such that
the Higgs field always remains massless.
However, there is no principle by which this
scale-invariance or masslessness of the Higgs field
can be maintained. After all, one is invoking
large Yukawa couplings, which feed
the notorious divergent quadratic corrections
to the Higgs boson mass.
The one-loop two-point and four-point functions
with quark $Q$ in the loop would generate
effective $\mu^2$ and $\lambda$ self-coupling
terms for the Higgs field.
With no explicit dynamical principle
(such as gauge invariance for the case of QED),
the assumption of a \emph{massless} Higgs doublet
as the agent of DSB is not only \emph{ad hoc},
but clearly unsustainable.

In contrast, we see the merit, as well as the meaning,
of our ``bootstrap" gap equation.
So long we are in the broken phase,
there is a massless Goldstone boson,
which couples with the known Yukawa coupling.
Treating the Yukawa coupling as large,
if a nontrivial solution to the gap equation is found
(as we have illustrated above),
it in turn justifies the use of a
massless Goldstone boson in the gap equation.
In fact, the physical argument~\cite{Hou12} was to
view the Goldstone boson $G$ as an extremely tight
ultrarelativistic bound state of heavy $Q$ and $\bar Q$
from the broken phase, while $G$ enters the
``bootstrap" gap equation to dynamically generate $m_Q$
hence break the symmetry, and in same stroke
justify its own existence.

From the argument above, we see that our bootstrap gap equation
puts the existence of the physical Higgs boson in doubt.
After all, it is rather difficult to keep a light physical Higgs boson,
in the presence of much stronger Yukawa couplings than the top quark.
For our gap equation, one also cannot ignore corrections to the Higgs propagator.
At the foundation level~\cite{Hou12},
unlike the Goldstone boson,
even though a 126 GeV Higgs-like boson has appeared,
its true nature has to be settled by experiment.
Our numerical study also shows that keeping the Higgs term
tends to raise the critical Yukawa coupling considerably,
implying $2m_Q$ higher than the LHC collision energy.
Thus, in the name of simplicity and pertinence,
we drop the $\kappa_b = \kappa_a$ case from now on.
However, we shall return at the end to discuss the dilaton issue.


\subsubsection{ Physical Cutoff of  $2m_Q < \Lambda_{\rm NP}$ }

The previous comparison with Hung-Xiong approach,
and in particular the bootstrap nature of the
Goldstone boson $G$ in the gap equation, illustrates that
the limit $\Lambda_{\rm UV} \to \infty$ would take us
outside the range of validity of the gap equation itself.
It is clear that, for timelike $p^2 > (2m_Q)^2$,
there is no Goldstone boson.
Thus, for the cutoff $\Lambda$, one should not use
the traditional language of $\Lambda_{\rm UV}$,
and one probably should not contemplate ``UV-completion"
at the current stage.
The bootstrap gap equation does not provide a \emph{theory} of
the heavy quark Yukawa coupling $\lambda_Q$,
but employs it for DSB.
Thus, we suggest a cutoff $\Lambda < 2m_Q < \Lambda_{\rm NP}$,
where $\Lambda_{\rm NP}$ is some true New Physics
scale where the origin of Yukawa couplings may be
contemplated, but it is out of reach for now.

Another possible interpretation of this cutoff is
related to the restoration of symmetry.
The Goldstone boson couples to broken currents,
but the $Z$ factor of the Goldstone boson
may vanish at some scale related to the scale
where the Goldstone boson becomes unbounded,
and symmetry is restored.
Rather than a true $\Lambda_{\rm UV}$ that
in principle extends to infinity,
there exists a cutoff $\Lambda$ of the gap equation.
We are in effect summing over the Goldstone boson
correction to the self-energy of the quark $Q$,
when the Goldstone boson is still defined.
As noted in Ref.~\cite{Hou12}, this picture does
receive experimental support, in that no
New Physics seem to be there below $\sim $ TeV scale.
So, one sums up only the effect of the Goldstone boson,
and nothing else.

\begin{figure}[t!]
\begin{center}
\vskip0.4cm
 \includegraphics[width=60mm]{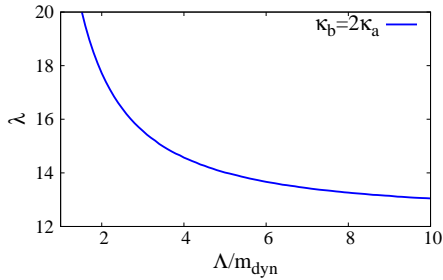}
\vskip0.3cm
\caption{Cutoff dependence of Yukawa coupling,
         where the cutoff $\Lambda$ is in units
         of the dynamical quark mass $m_{\rm dyn}$.}
 \label{lamLam}
\end{center}
\end{figure}

The gap equation sums up the effect,
from low to high momentum, of the correction
by the Goldstone boson $G$ to the quark self-energy.
Eq.~(\ref{lamcr_infty}) reflects taking this
sum to $\Lambda_{\rm UV} \to \infty$.
Since the summation is accumulative,
if now one sums only to some $\Lambda \lesssim 2m_Q$,
as we have argued, then the cumulative effect
is less than summing to very large momentum above $2m_Q$.
We note that $2m_Q$ is a physical scale parameter
that is external to the scale-invariant gap equation.
Nevertheless, we can plot the dependence of
$\lambda_Q^c$ on the cutoff $\Lambda$.
From Fig.~\ref{lamLam}, one can see that, for a lower cutoff,
$\lambda_Q^c$ has to be higher than in Eq.~(\ref{lamcr_infty}).
This is because, as the integration range is smaller,
a larger $\lambda_Q$ value is needed to compensate.
Thus, Eq.~(\ref{lamcr_infty}) gives a lower bound on $m_Q$.
If we take $\Lambda = 2m_Q$, then $\lambda_Q^c \sim 17.7$,
and
\begin{equation}
 m_Q \sim 3\ {\rm TeV}, \quad\quad {\rm (No\ Higgs;\ } \Lambda = 2m_Q)
 \label{mQcr_2mQ}
\end{equation}
which is very high.
We will return to discuss whether this could be
an overestimate later.



\subsubsection{ Decay Constant and Yukawa Coupling }

Up to now, we have been cavalier in the relation
between $\lambda_Q^c$ and $m_Q^{(c)}$,
treating it as $\sqrt2m_Q^{(c)} = \lambda_Q^c v$.
But $v$ is a physically measured value, and
$m_Q$ is yet to be experimentally measured.
If electroweak symmetry is indeed dynamically broken
by the large Yukawa coupling of a new heavy chiral quark $Q$,
when $Q$ is discovered in the future, likely $m_Q > m_Q^c$.
We then see that the actual v.e.v. value, $v$,
may not correspond to the critical value $\lambda_Q^c$.
This brings about the question of how scale invariance
is actually broken in our gap equations,
and whether there might be a dilaton~\cite{dilaton2}.
This is an extremely interesting question, given the possibility,
needed for the viability of our gap equation,
that the observed 126 GeV Higgs-like boson
might actually be a dilaton.
Our framework, however, does not provide the means to approach
this problem, as it is empirical and shuns the UV.

Rather than approach this deeper problem,
we try to obtain the decay constant $f_G$
of the Goldstone boson, which should be
the same as the vacuum expectation value, $v$.
Following the Pagels--Stokar formula~\cite{Pagels79}
naively, we obtain,
\begin{equation}
f_G^2 =
\frac{N_C}{4\pi^2} \int_0^{\Lambda^2}dx \frac{x B(x)^2 - \frac14 x (B(x)^2)^\prime}{(x A(x)^2+B(x)^2)^2}.
\end{equation}
%
More generally, we write
\begin{eqnarray}
f_G^2 &=& \int_0^{\Lambda^2} dx {\cal F}[ A(x),B(x) ] \nonumber \\
 &=& m_Q^2 \int_0^{\Lambda^2/m_Q^2} d\hat x \hat{\cal F} [ A(\hat x), B(\hat x)/A(\hat x)/m_Q ],
\end{eqnarray}
where $m_Q \equiv B(0)/A(0)$, and we have scaled by $m_Q$
(and redefined the function ${\cal F}$), treating it as physical.

We can get back the ``Yukawa" coupling $\lambda$
(it should really be denoted as $Y$,
and the question is whether $Y = \lambda$)
for the input to the gap equation, or
\begin{equation}
\frac1{2\lambda^2} = \int_0^{\Lambda^2/m_Q^2} d\hat x
                      \hat{\cal F} [ A(\hat x), B(\hat x)/A(\hat x)/m_Q ].
 \label{Y}
\end{equation}
If the system is really scale-invariant, the r.h.s. of
Eq.~(\ref{Y}) is a function of $\Lambda/m_Q$ and $\lambda$.
In order to satisfy the gap equation,
$\lambda$ is obtained as a function of $\Lambda/m_Q$.
Namely, it should not depend on $m_Q$ explicitly.
Taking the cutoff $\Lambda = 2 m_Q$, the equation becomes
iterative for $\lambda$.
Therefore, solving the gap equation,
we obtain a prediction for the heavy quark mass.
But technical issues remain:
Is $m_Q$ a physical mass? What about the infrared cutoff?
Can our assumption of
$\lambda (\Lambda_{\rm UV}/m_Q) =Y(\Lambda_{\rm UV}/m_Q)$
be maintained self-consistently?
We leave these theoretical questions to a future work.


\section{ Discussion }


\noindent{\ \emph{$\bullet$ Can $m_Q$ be brought lower than 2--3 TeV?}}
\vskip0.1cm

From a phenomenlogical standpoint, our numerical value of
$m_Q \sim 3$ TeV of Eq.~(\ref{mQcr_2mQ}), even the
2 TeV value of Eq.~(\ref{mQcr_infty}), seem depressingly high.
We offer a few remarks how this might be lowered.
In the spirit of Refs.~\cite{Hou12} and \cite{EHY11},
the Goldstone boson $G$ is the lowest or
most tightly bound state through the Yukawa coupling itself.
Yukawa-bound resonances exist above this
isotriplet, color-singlet, pseudoscalar state,
the leading ones being the
isotriplet, pseudoscalar $\pi_8$,
the isosinglet, vector $\omega_8$, which are both color-octet,
and the isosinglet, vector, color-singlet $\omega_1$ ``mesons".
Without solving the strongly coupled bound state problem,
one does not know the spectrum (i.e. how tightly they
are bound below $2m_Q$), nor their ``decay constants",
i.e. how they couple to the heavy quark $Q$.
But the couplings should be rather strong.
The point is, as one integrates the Goldstone loop
up to $2m_Q$, at some point these heavy mesons should
also enter, and contribute in the same spirit to
the self-energy of $Q$.

There is thus some hope that extra, attractive contributions
could lower $\lambda_Q^c$.
But it also illustrates the limits of our bootstrap approach.
The momentum integration for these extra contributions
start from the meson mass, up to $2m_Q$,
but clearly the meson propagators and the meson-$QQ$ vertices
would be much more sensitive to $q$ as it varies,
compared to the Goldstone boson $G$.
Even for $G$ (which is the $\pi_1$), as one approaches $2m_Q$,
its bound state nature would lead to modifications
of its propagator (even if symmetry remains broken
hence it remains massless) and vertex.

\vskip0.2cm
\noindent{\ \emph{$\bullet$ Self-consistency with Dilaton}}
\vskip0.1cm

We have already offered our critique of the work of Hung and Xiong~\cite{HX11},
and showed also numerically that the needed $\lambda_Q^c$ is
exorbitantly high if one includes a light SM Higgs.
Thus, if the new 126 GeV boson is found to be truly
the SM Higgs, our bootstrap DSB gap equation cannot work.
However, we have argued that the nature of the new boson
probably cannot be demonstrated beyond doubt with 2011-2012 LHC data,
and a loophole is the dilaton~\cite{dilaton, dilaton2}.
The dilaton is allowed by our gap equation,
since the DSB also breaks scale invariance.

At the operational level, one can easily see that,
if the observed 126 GeV object is a dilaton, it does not
change the main result of our study.
That is, given that the dilaton coupling is suppressed by $v/f$
compared to the SM Higgs boson,
where $f$ is the dilaton decay constant,
the $m_h$-dependent terms of Eqs.~(\ref{Bp2}) and (\ref{p2Ap2})
are suppressed by $v^2/f^2$,
and can be treated as subdominant hence dropped,
in the same spirit that the strong and weak gauge couplings
are treated as subdominant.


\vskip0.2cm
\noindent{\ \emph{$\bullet$ Comparison with NJL-type Models}}
\vskip0.1cm

A different question is whether our gap equation is
actually equivalent to the NJL model.
We have already commented that for the NJL model,
the self-energy does not depend on momentum,
and one simply cuts the loop momentum off at some $\Lambda$.
For our gap equation, the wave function part, $A$,
has momentum dependence, i.e. the Yukawa loop always
modifies the $Z$ factor.
We already saw this in scale-invariant QED.
For NJL, the cutoff $\Lambda$ is traded, together
with the associated dimension $-2$ coupling constant $G$,
for the physical $f_\pi$ and $m_N$, although,
depending on the cutoff, there is a
critical coupling $G_{\rm crit}$ (see Eq.~(\ref{Gcrit})).
For our case, one cannot take arbitrary values for the cutoff.
Instead, we argued that, because the Goldstone boson
would become unbounded at some scale, say $2m_Q$,
the cutoff of the loop momentum has to be ``heuristically" finite.
Further similarities and differences are noted
in Ref.~\cite{Hou12}.
A fundamental difference may be that one has
effectively postulated that the
dimension zero Yukawa coupling of the Goldstone boson
to be the experimentally verified one related to the
left-handed vector gauge coupling of massive quarks.
If the Goldstone boson $G$ is an ultratight
$Q\bar Q$ bound state, it has turned the
effective $Q\bar Q$ dimensionality to 1.
In this sense, our gap equation may resemble
the gauged NJL model \cite{dilaton1},
in which the dimensionality of the bound state
tends to 1 near the critical gauge coupling.
%
%

Our approach is also conceptually different from
those descended from the top condensation model~\cite{top_condense},
in which the gauged NJL is applied to EWSB.
The self-energy in the top condensation model incorporates
both our Fig.~2 (loop with four-quark operator) and
Fig.~3 (but with gauge boson in the loop).
The gap equation with the four-quark operator
is equivalent to the minimization of the linear sigma model
with a compositeness condition,
such that the Yukawa and Higgs quartic couplings
blow up at the composite scale.
Naively speaking, our gap equation of Fig.~4 corresponds to
the linear sigma model with large Yukawa coupling.
One can therefore read off the schematic
correspondence between top condensation and our approach
by replacing gauge coupling with Yukawa coupling,
and four-quark operator loop by possible heavy bound state loop
that we have discussed earlier.
In the top condensation model, the four-quark interaction
generates the large Yukawa coupling,
and thus, it is clear that it differs from our approach.
We did not touch the origins of Yukawa couplings at all.


\vskip0.2cm
\noindent{\ \emph{$\bullet$ Phenomenology at LHC?}}
\vskip0.1cm

Can our high quark mass of a couple of TeV,
with the associated ultra strong Yukawa coupling,
be testable at the LHC?
First, as we mentioned already, with this setup,
it is very hard to believe that there would be
a light Higgs at 126 GeV,
hence the observed new boson ought to be a dilaton~\cite{dilaton,dilaton2},
which can be checked experimentally.
Furthermore, one expects a heavy Higss $m_H > 600$ GeV,
and the search for heavy Higgs boson should continue to be pursued.

But even though there may be scalar resonances,
experience from hadronic physics (the $\sigma$ resonance)
suggest that this path may be rather murky.
So, besides the current search approach that
assumes  QCD pair production of $Q\bar Q$,
followed by free quark $Q$ decay,
what else might one do at the LHC?
After all, 2 or 3 TeV heavy quarks are approaching the searchable
limits even for the high luminosity LHC running at 14 TeV.
It was pointed out recently by one of us~\cite{Hou12-2},
making analogy with $\bar p p \to n\pi$ annihilation,
that the search strategy should contain
$\bar QQ \to nV_L$, where $V_L$ is nothing but
the Goldstone bosons $G$ of the electroweak sector.
It was argued that the multiplicity would be
high, and behave as thermal emission from a
``fireball" with temperature $T$ that is related
to the v.e.v. scale $v$.
If the bound states, especially $\omega_8$,
decay in a similar way via multi-$V_L$, then there is
good hope for such spectacular phenomena at the LHC.
A corollary~\cite{Hou12-2} is that the $V_LV_L \to V_LV_L$ channel
may be the wrong path for going beyond
heavy Higgs search, but it rather should be
$V_LV_L \to nV_L$.
What is intriguing is that the physically measured
$g_{\pi NN}$ coupling is consistent with
$\lambda_{\pi NN} \equiv \sqrt2 m_N/f_\pi$,
and is rather similar in value to Eq.~(\ref{lamcr_infty}).
This offers a totally separate argument, hence giving
confidence, that $m_Q$ is in fact of order 2 TeV or above.

\vskip0.2cm
\noindent{\ \emph{$\bullet$ Revisiting Fermi--Yang Model}}
\vskip0.1cm

Having made the analogy of $G$-$Q$ with $\pi$-$N$,
it brings back the question of
whether the pion could really have been
an $N\bar N$ bound state, i.e. whether
the original conjecture of Fermi and Yang~\cite{FY49},
could have been realized.
The strength of $\pi NN$ coupling is simply staggering.
But subsequent developments in hadron physics experiment
relatively quickly gave rise to meson states
in 500 to 800 MeV range (and corresponding
baryon resonances), eventually exploding,
in 1 to 2 GeV range, i.e. below $2m_N$.
Thus, our gap equation does not apply.
Even if one employed our gap equation,
one cannot integrate the pion loop up to $2m_N$:
the integral is cutoff at a scale $\sim \Lambda_{\rm QCD}$
that is quite below $2m_N$.
And it turned out that $\pi$ and $N$ were QCD bound states of fermi size,
so the pion was not an ultratight bound state of $N$ and $\bar N$.

But the possible existence of a very heavy chiral doublet
offers us another chance.
If the $Q\bar Q$ ultratight bound state picture for
the Goldstone $G$ could be realized according to
our bootstrap gap equation, it would
strengthen our reasoning~\cite{Hou12}
that the underlying theory for Yukawa couplings
cannot be a simple mock-up of QCD, such as
(the various forms of) technicolor (TC).

We do not have new insight on bound state phenomena,
other than what is already discussed in Ref.~\cite{EHY11}.
Unfortunately, this reference was very conservative
and did not discuss above $m_Q > 700$ GeV,
as the Bethe--Salpeter (BS) equation approach
tend to have collapsed states.
But this was, in turn, the foundation for the postulate
made in Ref.~\cite{Hou12} that the leading collapsed state,
$\pi_1$, is precisely the Goldstone boson $G$,
which lead to the present gap equation study.

We do not yet know whether our gap equation could
shed any light on the $Q\bar Q$ bound state spectra.
The SD equation itself is of course
a ``higher level" one than the BS equation for bound state.
If the BS equation Yukawa-boundstate approach
can be a guide for $m_Q$ as high as 2 TeV,
i.e. with $\alpha_Q > 1$, the noteworthy point is that
the leading bound states~\cite{EHY11} $\omega_1$, $\omega_8$ and $\pi_8$
are rather distinct from the $\rho$s and $\eta$s of TC.
Since $\rho$-like states are the typical working assumption
for DSB that tends to adopt the QCD or TC mindset,
we wish to stress this distinguishing aspect of Yukawa-induced DSB.

\vskip0.2cm
\noindent{\ \emph{$\bullet$ Question of Flavor vs EWSB}}
\vskip0.1cm

Bound states in our approach emerge from strong Yukawa coupling
(rather than QCD-like gauge dynamics as in TC),
but we did not offer any theory of Yukawa couplings.
This may be an advantage:
By simply employing Yukawa couplings,
we inherit this well tested part of the SM,
including the flavor sector.

The three generation SM can account for all observed
flavor and CP violation phenomena, with only minor tensions
after a decade of detailed scrutiny by the B factories.
Note that the Higgs boson does not enter flavor processes
of interest, such as box and electroweak penguin diagrams;
all interesting effects arise from Goldstone couplings.
Extending to a 4th generation to address EWSB by
strong Yukawa coupling, we will not encounter the
usual issues of ``flavor scale" as in most other approaches.
Mixing of the 4th generation with lower ones must be suppressed,
as the LHCb experiment finds all key measurements,
such as CP phase in $B_s  \to J/\psi \phi$,
$A_{\rm FB}$ in $B\to K^*\mu\mu$, as well as $B_s \to \mu\mu$
are all consistent with SM.
But this follows the well known pattern that CKM matrix elements
trickle down in strength as one goes far off-diagonal.
We do note that, having a 4th generation could seemingly
provide enough CP violation strength~\cite{HouCPV4}
for matter dominance of the Universe,
which can be viewed as an independent motivation
for continuing to entertain the 4th generation.

Thus, unlike most approaches that suffer the dilemma of
need for TeV scale physics to ``stabilize the Higgs" (or EWSB)
on one hand, while having a much higher flavor scale,
in our case, the flavor physics scale need not be that far off.
If realized, the actual origins of Yukawa couplings
would become the focus question.

%
We mention in passing that
our gap equation can be easily extended to finite temperature,
allowing one to potentially explore issues
related to electroweak phase transition,
which is a direction that we would take up
in a subsequent work.

\section{Conclusion}

Despite the emergence of a Higgs-like new boson with mass of order 126 GeV at the LHC,
we have inspected current data up to 17--18 fb$^{-1}$ level, and
concluded that a dilaton interpretation cannot be fully ruled out,
even if it appears fortuitous.
Assuming this state as a dilaton that feigns (for now) the SM Higgs boson,
while the actual Higgs boson is above 600 GeV and heavy,
we consider seriously the possibility of 
electroweak symmetry breaking driven by strong Yukawa coupling.

Starting from a purely empirical basis, a dynamical gap equation 
is argued, treating the Goldstone as massless inside the loop,
coupling with chiral quark doublet $Q$ with the usual Yukawa couplings.
By empirical we mean the traditional sense of based on experimentally established facts,
where we have utilized electroweak gauge symmetry and its
spontaneously broken nature, only extending by a new chiral doublet $Q$
 --- the 4th generation --- with Yukawa coupling $\lambda_Q$ 
already above the nominal unitarity bound, which is again empirical.
The gap equation effectively sums over exchange momentum of
$Q\bar Q$ scattering.
We further utilize the experimental fact that there are no
obvious new states below 1 or 2 TeV, hence this integration range
can extend up to ``$2m_Q$" without any further significant contributions, 
while $m_Q$ is to be determined by solving this dynamical equation.
This is done by drawing experience from strongly coupled, 
massless (hence scale invariant) QED.
Unlike QED, where choosing the Landau gauge leads to simplification,
in the present case one needs to face a coupled integral equation.
Numerical solutions are found, hence dynamical EWSB demonstrated,
at the cost of staggeringly high quark mass $m_Q$ in the 2--3 TeV range.
Though rather high in value, LHC might still shed light on it,
as the critical $\lambda_Q \gtrsim 4\pi$ turns out analogous
to the $\pi$--$N$ system, which provides some justification.
It also suggests, by analogy, the possible novel phenomenon of 
multi-$G$, or multi-$V_L$ production at 14 TeV LHC.

We have already stated our preference for the heavy or 
``no Higgs" (not the same as Higgsless) scenario.
However, in its stead the existence of $Q\bar Q$ mesons in the form of
color octet $\pi_8$ and $\omega_8$, and color singlet $\omega_1$
are implied, where the notation is under a heavy isospin $I_Q$.
It is interesting that a dilaton is \emph{allowed} by our gap equation, 
which is nominally scale invariant, while our
dynamical electroweak symmetry breaking solution
also breaks this scale invariance. But the actual source of this
scale invariance violation is likely rooted in the 
dynamical origins, at a considerably higher UV scale, of $\lambda_Q$, 
which was only treated as a parameter in our present work.
The LHC could establish the dilaton nature of the 126 GeV boson,
if it is confirmed that vector boson fusion (VBF) and Higgsstralung (VH)
production processes are indeed suppressed.
Measuring this suppression factor would tell us the dilaton scale, 
or decay constant $f$, while the 126 GeV mass would be 
a ``messenger" from higher UV theory of the actual scale invariance violation.
Thus, if a dilaton emerges, and our dynamical equation is
confirmed in some form, it is certainly no less exciting
than the discovery of the SM Higgs itself.

\vskip0.3cm
\noindent{\bf Acknowledgement}.
We thank H.-C. Cheng, P.Q.~Hung, T~Kugo, C.N. Leung and M. Piai for discussions.
The research of YM is supported by NTU grant NTU-98R0526,
and HK by National Science Council grant NSC-99-2811-M-033-017 of Taiwan,
and the National Research Foundation of Korea funded by the
Korean Government (Grant No. NRF-2011-220-C00011).
WSH is supported by NSC 100-2745-M-002-002-ASP
and various NTU grants under the MOE Excellence program.


\begin{thebibliography}{9}
\bibitem{Tev}
  There is indication from the Tevatron for $VH$ production
  in the $H \to b\bar b$ channel. See the plenary talk by
  Aurelio Juste at HCP 2012 for the latest update.
\bibitem{foot}
  We remark that, physically speaking, our gap equation cannot
  be analogous to strong QED, which is a vector theory, while
  we have pseudoscalar particle in the loop.
\end{thebibliography}
\end{document}